\documentclass{ws-procs975x65}

\begin{document}

\title{Theoretical interpretation of GRB060124: preliminary results.}

\author{R. Guida$^*$, M.G. Bernardini, C.L. Bianco, L. Caito, M.G. Dainotti and R. Ruffini}

\address{Dipartimento di Fisica, Universit\`a La Sapienza,\\
Roma, 00185, Italy\\
$^*$E-mail: roberto.guida@icra.it\\
www.icra.it}

\begin{abstract}
We show the preliminary results of the application of our ``fireshell'' model to GRB060124. This source is very peculiar because it is the first event for which both the prompt and the afterglow emission were observed simultaneously by the three \emph{Swift} instruments: BAT ($15-350$ keV), XRT ($0.2-10$ keV) and UVOT ($170-650$ nm), due to the presence of a precursor $\sim 570$ s before the main burst. We analyze GRB060124 within our ``canonical'' GRB scenario, identifying the precursor with the P-GRB and the prompt emission with the afterglow peak emission. In this way we reproduce correctly the energetics of both these two components. We reproduce also the observed time delay between the precursor (P-GRB) and the main burst. The effect of such a time delay in our model will be discussed.
\end{abstract}


\bodymatter

\section{GRB060124 observational properties}
On 2006-01-24 at 15:54:52 UT, Swift-BAT triggered on the precursor of GRB060124, that occurred $\sim 570$\,s before the main burst peak \cite{romano-060124}. This allowed Swift to immediately re-point the narrow field instruments (NFIs) and acquire a pointing towards the burst $\sim 350$\,s {\it before} the main burst occurred. The burst has a highly structured profile, comprising three major peaks following the precursor and has the longest duration (even excluding the precursor) ever recorded \cite{laz05}.

GRB060124 also triggered Konus-Wind ($10-770$ keV) \cite{KonusWind} 559.4 s after the BAT trigger \cite{golenetskii2006:gcn4599}. The Konus light curve confirmed the presence of both the precursor and the three peaks of prompt emission.

The  prompt emission of GRB~060124 was observed simultaneously by XRT with exceptional signal-to-noise (S/N) and was detected by UVOT at $V=16.96\pm 0.08$ ($T+183$\,s) and $V=16.79\pm 0.04$ ($T+633$\,s) \cite{romano-060124}. This fact makes it an exceptional test case to study prompt emission models, since this is the very first case that the burst could be observed with an X-ray CCD with high spatial resolution imaging down to $0.2$ keV.

\section{The fit}
Within our ``canonical GRB'' scenario \cite{XIIBSCG} we identify the first main pulse with the P-GRB and the three major peaks following the precursor with the afterglow peak emission.

We therefore obtain for the two parameters characterizing the source in our model $E_{e^\pm}^{tot}=3.73\times 10^{54}$ erg and $B = 2.3\times 10^{-3}$. This implies an initial $e^\pm$ plasma created between the radii $r_1 = 1.12\times10^7$ cm and $r_2 = 4.58\times10^8$ cm with a total number of $e^{\pm}$ pairs $N_{e^\pm} = 1.46\times 10^{59}$ and an initial temperature $T = 2.23$ MeV.

The theoretically estimated total isotropic energy emitted in the P-GRB is $E_{P-GRB}=1.41\% E_{e^\pm}^{tot}=5.26 \times 10^{52}$ erg, in excellent agreement with the one observed in the first main pulse ($E_{P-GRB}^{obs} \sim 6.00 \times 10^{52}$ erg in $15-350$ keV energy band), as expected due to their identification. After the transparency point at $r_0 = 4.76\times 10^{14}$ cm from the progenitor, the initial Lorentz gamma factor of the fireshell is $\gamma_0 = 430$. The distribution of the CircumBurst medium has been parametrized assuming an average value for the effective density in the prompt phase of $10^{-2}$ particle per cm$^{3}$ and in the afterglow phase of $10^{-4}$ particle per cm$^{3}$. Such a low effective density has been assumed in order to reproduce the $\sim 500$ s of quiescence between the P-GRB and the prompt, according to the way in which the emission is produced within our model, that it will be clarified in the next session.

\begin{figure}[t]
\begin{center}
\psfig{file=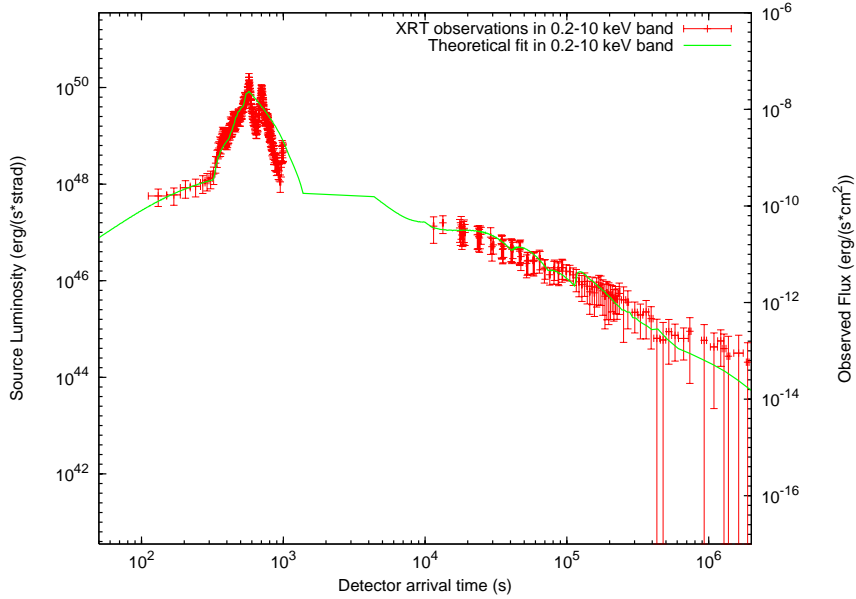,width=0.9\hsize}
\end{center}
\caption{The XRT light curve ($0.2$--$10$ keV, red points) and the preliminary theoretical simulation in the same energy band (green line). The fit is quite good, but the double peaked structure is not reproduced, due to the fact that our radial approximation for modeling the CBM is not valid anymore at the late time of the peaks (see text).}
\label{xrt}
\end{figure}

\begin{figure}[t]
\begin{center}
\psfig{file=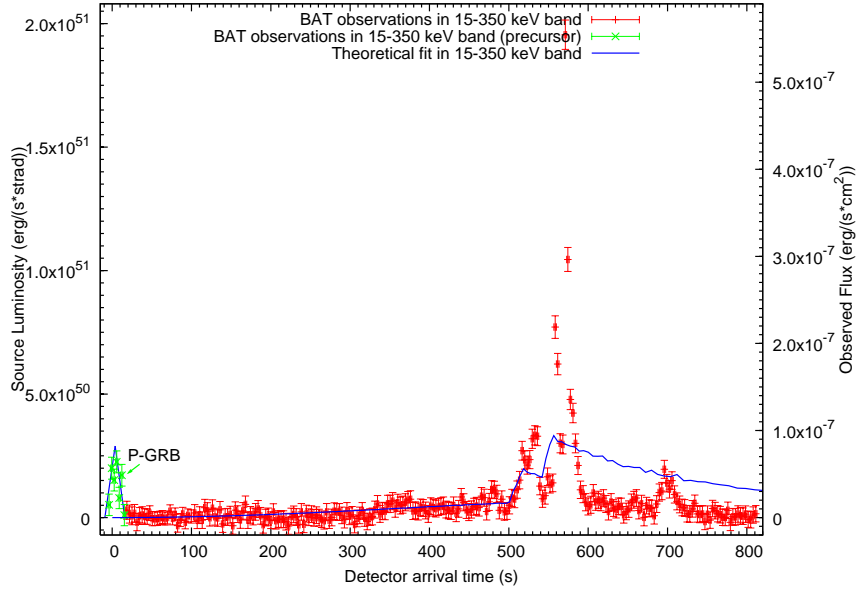,width=0.9\hsize}
\end{center}
\caption{The BAT light curve in the $15-350$ keV band (red points) comprising also the precursor (green points) and our preliminary theoretical simulation in the same energy band (blue line). Clearly the energetics is well reproduced, but in order to have a good fit of the peaks, a correct treatment of the 3-dimensional structure of the CBM is needed (see text).}
\label{bat}
\end{figure}

In Fig.~\ref{xrt} we present the preliminary theoretical fit of the \emph{Swift} XRT data ($0.2$--$10$ keV), while in Fig.~\ref{bat} of the BAT ones ($15$--$350$ keV). The problems of the fit will be discussed in the next section.

\section{The CircumBurst 3D structure}

Within our fireshell model all the GRB emission after the transparency is produced by the interaction of the accelerated baryons with the CBM, and such interaction is modeled as inelastic collisions \cite{ruf02}. The number of such collisions, hence, depends on the CBM density.

The simplest way to model the CBM structure is to assume that $n_{cbm}$ is a function only of the radial coordinate, $n_{cbm}=n_{cbm}(r)$ (radial approximation). The CBM is arranged in spherical shells of width $\sim 10^{15}$ cm positioned in such a way that the modulation of the emitted flux coincides with the observed peaks. It is important to emphasize that, when the accelerated baryons collide with a shell, the increase in the flux is almost immediate due to the photons coming from the line of sight. Then it follows an exponential decrease of the flux due to the contribution of the photons emitted from different angles. In this way we obtain the observed FRED structure for each peak, together with all the other observed peculiarities (hard to soft transition, spectral lag).

Clearly our radial approximation is valid until the visible area of the incoming baryons pulse is comparable with the characteristic dimensions of the clouds. The transverse dimension of such area is $R_T=r \sin \theta$, where $\theta \sim 1/\gamma$ is the relativistic beaming angle, so we have $R_T \sim r/\gamma$. 

We have found in many cases that this approximation cannot be valid during the whole prompt emission. In fact, when the accelerated baryons impact with dense clouds of CBM, they are decelerated and their gamma factor drops abruptly. In this situation, after the first peaks (the number of peaks depending from their height, the higher they are the smaller their number is) the visible area becomes comparable with the size of the clouds and our approximation is not valid anymore. This is case for other GRBs we analyzed, as GRB991216 \cite{ruf02} and GRB050315 \cite{ruf06a} .

Another situation in which our radial approximation fails can occurs. Because the transverse dimension of the baryonic fireshell's visible area, as outlined above, depends not only from the Lorentz gamma factor but also from the radius of the fireshell, it can be that for very large value of this radial coordinate, the size of the visible area becomes comparable with the CBM clouds, that is, the approximation of spherical symmetric distribution for the CBM fails.

In all the GRB sources studied up to date, this have never been the case, because usually the radial coordinate $r$ at which the prompt emission occurs is small.

It is important here to remember the fact that within our fireshell model, the initial instant of time $t_0$ (related to the initial value of the radial coordinate, $r_0=c t_0$) is often different from the moment in which the satellite instrument triggers: in fact in our model the GRB emission starts at the transparency point when the P-GRB is emitted, but sometimes the P-GRB is under the instrumental threshold or comparable with it and so is not enough to trigger the instrument. For example in the case of GRB050315, a possible precursor was observed $\sim 50$ s before the trigger \cite{vau06}, that indeed occurred when the main prompt emission started. 

In this case instead the BAT instrument triggers on a precursor that we identify as the P-GRB because of the excellent agreement in terms of the energetics and of the time delay between it and the main prompt emission; so in this case our $t_0$ coincides with the BAT trigger and the main prompt emission occurs at $\Delta T\sim 600$ s so at a value $r=c \Delta T$ for the radial coordinate of the fireshell; with this value of $r$ the transverse dimension of the baryonic fireshell's visible area is such that the radial approximation is not valid anymore.

In particular, we found that at $t_a^d\sim 600$ s, that is when the main burst occurs, the radius of the fireshell is $r\sim 10^{18}$ cm and the Lorentz gamma factor has dropped abruptly to a value of $\sim 100$ from the initial $\gamma_0 = 430$, due to the CBM cloud assumed to be present at the moment of the prompt emission. The transverse dimension of the visible area of the incoming baryons pulse indeed results $R_T\sim 10^{16}$ cm, so even bigger then the characteristic dimensions of the CBM clouds usually assumed (from \cite{ruf02} $10^{14}$ to $10^{15}$ cm), in this case $\sim 10^{15}$ cm.

A correct treatment of the 3-dimensional structure of the CBM clouds is needed in this case.

We have already tested this idea in order to explain an apparently physical different feature of the GRBs: the flares. This phenomenon has been discovered to occurs in the early part of the X-ray afterglow, that means very late from the satellite trigger and very far. From our point of view, there are no differences between a flare and the prompt emission in this case, that has occurred at $600$ s.

Many interpretations have been provided in order to explain the flares. The most common explanation is a central engine activity which results in internal shocks (or similar energy dissipation events) at later times \cite{zhang} . Another possibility is emission from reverse shock, but the predicted amplitude is too low to interpret all the cases \cite{Burrows05,zhang} . Alternatively such emission could be produced by a multi-component jet \cite{MR01,R-RCR02,KPi00a} : the X-ray flare is caused by the deceleration of the wider cocoon component with the ambient medium. In this case, however, the decay after the peak should follow the standard afterglow model, so it cannot interpret the observed rapid fall-off in the flares \cite{zhang} . The same problem \cite{zhang} affects also the scenario in which the flare is produced by the energy injection into the decelerating shell by the collision with a high-$\gamma$ shell \cite{KPi00b} .

Within our fireshell model the flares are interpreted as being due to the same process responsible for the following afterglow emission. So the difficulties to fit them are due to the radial approximation, not valid anymore at such late time (or at such big value of the radial coordinate).

We tested our idea of abandoning the radial approximation and introducing a 3-dimensional structure of the CBM clouds in order to fit the flare (occurred at $\sim 250$ s) of GRB011121 \cite{bia-cai-ruf,caia} , an old burst observed by \emph{Beppo}SAX which for the first time showed the feature of an X-ray flare. We obtain good results that demonstrate at least the validity of such proposal. Anyway the implementation of a such description of the CBM clouds is not yet finished, but we are currently working on it.

\section{Conclusion}
We applied the fireshell model to GRB060124. The work is not finished yet and we showed only the preliminary results. The main peculiarity of this source is the biggest ever recorded time delay between the precursor and the prompt emission. We reproduced correctly the energetics of the precursor, identified with the P-GRB, and of the prompt emission, identified with the extended afterglow peak emission. 

The most important consequence of having such a big time delay between P-GRB and afterglow peak is that the radial approximation assumed in modeling the CBM structure is not valid anymore at the time of the prompt emission. For this reason our model failed in reproducing the narrow two peaks of the prompt emission. Our peaks, in particular the second, resulted much more spread. 

In order to have a good fit of the light curves, we have to change our way of modeling the CBM structure. We have to take into account the fact that only a part of the visible area of the fireshell interacts with the CBM cloud. This is only possible introducing a 3-dimensional structure of the clouds, that will mean to introduce a new parameter. In this way we will obtain narrow peaks also for big values of the fireshell radius. 

We have already successfully applied this idea in order to fit the flare of GRB011121, that is a bump of an order of magnitude in luminosity, lasting for $20$ s, occurred after $250$ s from the trigger. The likeness of this flare with the prompt emission of GRB060124, a short bump of an order of magnitude in luminosity occurred at very late time as well, is evident: so we expect to obtain also in this case the same good agreement we had in the case of GRB011121.

\end{document}